\documentclass[12pt, draftclsnofoot, onecolumn]{IEEEtran}
\usepackage{graphicx}
\usepackage{amsmath,bbm,epsfig,amssymb,amsfonts, amstext, verbatim,amsopn,multirow,multicol,lipsum}

\usepackage[font=scriptsize]{caption}
\usepackage{subcaption}
\usepackage{balance}
\usepackage{url}
\usepackage{bbding}
\usepackage{amsfonts}
\usepackage{epsfig}
\usepackage{epstopdf}
\usepackage{setspace}
\usepackage{stmaryrd}
\usepackage{psfrag}	
\usepackage{multirow}
\usepackage{float}
\usepackage{cases}  
\usepackage[process=auto]{pstool}
\usepackage{etoolbox}
\usepackage{algorithm}
\usepackage{algorithmic}
\usepackage{hyperref}
\usepackage{wasysym}[nointegrals]
\usepackage{pifont}
\usepackage{enumitem}
\allowdisplaybreaks
\usepackage{graphicx}
\usepackage{wrapfig}
\usepackage{lscape}
\usepackage{rotating}
\usepackage{epstopdf}
\usepackage{rotating}


\newtheorem{theo}{Theorem}
\newtheorem{lem}{Lemma}

\newtheorem{prop}{Proposition}
\newtheorem{corol}{Corollary}

\newcommand{\fourie}[1]{\circ\!\!\!-\!\!\!\!-\!\!\!\!-\!\!\!\bullet}
\newcommand{\pushright}[1]{\ifmeasuring@#1\else\omit\hfill$\displaystyle#1$\fi\ignorespaces}
\newcommand{\pushleft}[1]{\ifmeasuring@#1\else\omit$\displaystyle#1$\hfill\fi\ignorespaces}
\newtoggle{Conf}

\togglefalse{Conf}

\iftoggle{Conf}{%
}{%
}

\EndPreamble
\usepackage{mathtools}

\begin{document}
	\title{  Delay Dispersion in  IRS-assisted FSO Links
		\vspace{-0.3cm}}
	\author{Hedieh Ajam\IEEEauthorrefmark{1},  Vahid Jamali \IEEEauthorrefmark{2}, Bernhard Schmauss\IEEEauthorrefmark{1}, and Robert Schober\IEEEauthorrefmark{1}\\  
		\IEEEauthorrefmark{1}Friedrich-Alexander-Universi\"at  Erlangen-N\"urnberg, \IEEEauthorrefmark{2}Technical University of Darmstadt
		\vspace{-0.4cm}}
	\maketitle
	\begin{abstract}
		The line-of-sight (LOS) requirement of free-space optical (FSO) systems can be relaxed by employing   optical intelligent reflecting surfaces (IRSs). 	 In this paper, we model the impact of  the IRS-induced delay dispersion  and derive  the channel impulse response (CIR) of IRS-assisted FSO links. The proposed model takes into  account the characteristics of the incident and reflected beams' wavefronts, the position of  transmitter and receiver, the size of the IRS, and the incident beamwidth on the IRS. Our simulation results reveal that a maximum effective   delay spread of 0.7 ns is expected for a square IRS with area 1 $\mathrm{m}^2$, which induces inter-symbol interference for bit rates larger than 10 Gbps. We show that the IRS-induced delay dispersion can be mitigated via equalization at the receiver. 
	\end{abstract}
	\section{Introduction}
	Free space optical (FSO) systems are prime candidates for  incorporation in next generation  wireless communication networks \cite{6G}. Due to their directional narrow laser beams, easy-to-install  transceivers, and high data rates,  FSO links are appealing for last-mile access, fiber backup, and backhaul of wireless networks.
	The line-of-sight (LOS) requirement of FSO systems can be relaxed by employing optical relays or optical intelligent reflecting surfaces (IRSs) \cite{Globecom_ScalingLaw}. Optical IRSs are planar structures   which can change the properties of the incident beam such as its phase and polarization \cite{Vahid_Magazine,Marzieh_IRS, TCOM_IRSFSO}.
	
	The authors in \cite{Marzieh_IRS_jou, Haas_IRS, IRS_FSO_WCNC} have developed  deterministic and statistical channel models for IRS-assisted FSO links. Based on these models, they  analyzed the impact of  IRSs on the normalized received power, known as geometric loss, and the misalignment loss due to random movements  of the  transmitter (Tx), IRS, and receiver (Rx). Moreover, in \cite{TCOM_IRSFSO, TCOM_IRSRelay}, the authors studied  the impact of the IRS size and  phase shift profile   on the received power at the Rx lens and compared the end-to-end performance with that achieved with optical relays. 
	
	Optical IRSs can be   implemented as planar surfaces equipped with arrays of micro-mirrors or meta-material-based passive subwavelength elements \cite{Vahid_IRS, AmplifyingRIS_Herald}. The former technology uses specular reflection by adjusting the orientation of the  micro-mirrors to redirect  the incident beam in the desired direction. 
	In contrast,  metamaterial-based IRSs change  the phase of the incident beam locally at every IRS element and provide anomalous reflection to point the beam towards the desired direction. Thus, for both types of IRSs, every IRS element reflects part of the beam incident  on the IRS. The parts of the beam reflected by different IRS elements have different  distances  to the Rx lens. Depending on the size of the beam footprint on the IRS,  the size of the IRS, and the position of  Tx and Rx, the different delays experienced by different parts of the reflected beam may cause IRS-induced channel dispersion.    
	Moreover, due to the high data rate of FSO links (typically on the order of 1-10 Gbps), the symbol duration is very small (0.1 - 1 ns). Thus, the delay dispersion  introduced by the IRS may cause  inter-symbol interference (ISI),  which degrades the end-to-end performance and needs to be mitigated. The authors in \cite{ISI_Zhang} investigated the impact of the time delay between different antenna elements of a phased array and the resulting ISI for wideband millimeter-wave signals. Moreover, the authors in \cite{ISI_Hass}  studied the frequency domain characteristics of IRS-assisted visible light communication (VLC) and the corresponding achievable rate. However, to the best of the authors' knowledge, the IRS-induced channel dispersion of FSO links has not been investigated in the literature, yet. 
	
	In this paper, we analyze the delay dispersion caused by IRS-assisted FSO links given the phase profile of the transmitted  and  reflected beams. Then,  we derive the channel impulse response (CIR) using the Huygens-Fresnel principle where we   take into account the size of the IRS and  the positions of  Tx and Rx. Furthermore, we mitigate the channel dispersion-induced ISI  using  digital equalization methods. Our simulation results reveal that the   delay spread   introduced by the IRS elements increases with  the difference between the angle of  the incident beam and the angle of the reflected  beam on the IRS, denoted by $\Delta\theta$.  Moreover, as $\Delta\theta$ increases, the channel gain decreases due to the higher geometric loss of the IRS-assisted FSO link.  Furthermore, we show that for  dispersive FSO channels  equalization  can greatly improve  performance.

	\section{System and Channel Models}\label{Sec_System}
\begin{figure}[t!]
	\centering
	\includegraphics[width=0.65\textwidth]{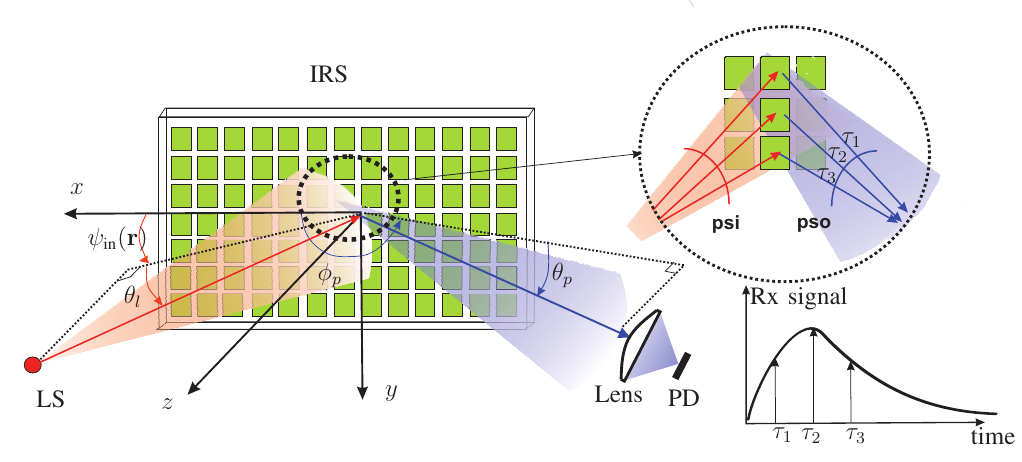}
	\caption{IRS-assisted FSO system model.}
	\label{Fig:System}\vspace{-0.8cm}
\end{figure} 
	
	We consider  a point-to-point FSO link,  where the Tx is equipped with a laser source (LS) and connected via an optical IRS to a  Rx, which is equipped with a photo detector (PD) and a lens, see Fig.~\ref{Fig:System}. The $xy$-plane  of the adopted $xyz$-coordinate system coincides with   the IRS plane and the  center of the IRS is in the origin.
	
	\subsection{System Model}
	
	\subsubsection{Transmitter}
	The Tx is equipped with an LS  emitting a Gaussian laser beam. The beam axis  intersects with the $xy$-plane at distance  $d_l$ and points in direction $\mathbf{\Psi}_{l}=\left(\theta_{l},\phi_{l}\right)$, where $\theta_l$ is the angle between the $xy$-plane and the beam axis, and $\phi_{l}$ is the angle between the projection of the beam axis on the $xy$-plane and the  $x$-axis, see Fig.~\ref{Fig:System}.
	
\begin{figure}[t!]
	\centering
		\includegraphics[width=0.85\textwidth]{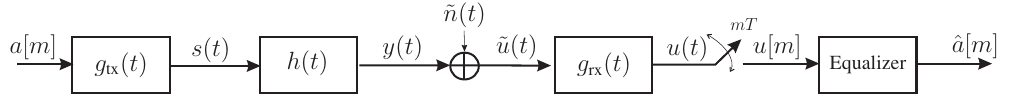}
	\caption{End-to-end optical system model.}
	\label{Fig:Equalizer}\vspace{-0.8cm}
\end{figure}

	We consider the transmission of a sequence  of independent and identically distributed (i.i.d.) symbols $a[m]$.  After  pulse shaping, the optical transmit  signal is given by 
	\begin{IEEEeqnarray}{rll}
		s(t)={P_0}\sum_{m=-\infty}^{+\infty} a[m] g_\text{tx}(t-mT),
		\label{}
	\end{IEEEeqnarray}
	where ${P_0}$ is the transmit power, $a[m]\in\{0,1\}$ are on-off keying (OOK) modulated symbols, and $g_\text{tx}(t)$ is the real-valued pulse shaping filter with symbol duration $T$, see Fig.~\ref{Fig:Equalizer}. 
	
	We assume  rectangular pulses, which are easy to realize in FSO systems \cite{Steve_DD} and are defined as
	\begin{IEEEeqnarray}{rll}
		g_\text{tx}(t)=\frac{1}{T}\Pi\left({\frac{t}{T}}\right), \quad  
		\label{TranmitPulse_TD}
	\end{IEEEeqnarray}
	with spectrum  $G_\text{tx}(f)=\int_{-\infty}^{\infty} g_\text{tx}(t) e^{-\mathrm{j}2\pi ft}\mathrm{d}t=\text{sinc}\left({T\pi f}\right)$. Here, $\text{sinc}(x)=\frac{\sin( x)}{ x}$ is the sinc-function and  $\Pi(\cdot)$  is defined as 
	$\Pi(\frac{x}{a})=1$ for $|x|<\frac{a}{2}$ and  $\Pi(\frac{x}{a})=0$ otherwise.
	

	\subsubsection{Receiver}
	The Rx is equipped with a PD and  a circular  lens of radius $a$.  The lens  of the Rx   is located at distance $d_{p}$ from the origin of the $xyz$-coordinate system. The normal vector of  the lens plane points in direction $\mathbf{\Psi}_{p}=\left(\theta_{p}, \phi_{p}\right)$, where $\theta_{p}$ is the angle between the $xy$-plane and the normal vector, and $\phi_{p}$ is the angle between the projection of the normal vector on the $xy$-plane and the $x$-axis.
	
	We assume a direct detection system, where the PD   produces an  electrical current, $y(t)$, which is proportional to the optical power received at the lens\footnote{We note that in  dispersive single-mode optical fibers, where the detector area is on the order of wavelength, the relation between the input optical signal and the output  of the dispersive optical channel is described by a nonlinear system \cite{Hanik_fiber}. However,  in FSO receivers, the lens  integrates the  electric field of the received optical signal over an area with a size of millions of square wavelengths, and focuses the beam on the PD which does not exploit the phase information of the electric field \cite{Kahn_experimental}. Thus,  the optical received power, $y(t)$, can be regarded as the received signal and the relationship  between transmit signal $s(t)$ and received signal $y(t)$ can be characterized by a linear system modeling the  dispersive IRS-assisted FSO link, c.f. (\ref{AfterPD}), \cite{Kahn_Infrared}.}, i.e., $y(t)=s(t)\circledast h(t)$, where  $\circledast$ denotes the convolution. Moreover, $h(t)$ is the CIR in equivalent baseband representation which will be discussed in detail in Section \ref{Sec_Channel}.  Assuming that the PD has no bandwidth constraint, after  filtering $y(t)$ with a receive filter $g_\text{rx}(t)$ matched to the transmit pulse, i.e., $g_\text{rx}(t)=g_\text{tx}^*(-t)$, the electrical signal $u(t)$ before sampling is given by
	\begin{IEEEeqnarray}{rll}
		u(t)&=\left({y}(t)+\tilde{n}(t)\right)\circledast g_\text{rx}(t)=r(t)+n(t),
		\label{AfterPD}
	\end{IEEEeqnarray}
	where ${r}(t)=y(t)\circledast g_\text{rx}(t)$  \cite{Kahn_Infrared, Steve_DD} and $(\cdot)^*$  denotes the complex conjugate  operator.    Furthermore, $\tilde{n}(t)$ is  additive white Gaussian noise with mean zero and power spectral density $N_0$. The filtered noise $n(t)=\tilde{n}(t)\circledast g_\text{rx}(t)$ is a zero-mean Gaussian process with autocorrelation function $\frac{N_0}{2}  \Phi_{{g_\text{rx}}{g_\text{rx}}}(\tau)$, where $\Phi_{{g_\text{rx}}{g_\text{rx}}}(\tau)=g_\text{rx}(\tau)\circledast g_\text{rx}^*(-\tau)$.


	
	\subsubsection{IRS}

	The size of the IRS is $L_{x}\times L_{y}$ and it consists of  $Q=Q_xQ_y$ subwavelength elements,  where $Q_x$ and $Q_y$ are the numbers of elements in  $x$- and $y$-direction, respectively.   Given that  the size of the IRS is much larger than the optical wavelength, i.e., $L_x, L_y \gg \lambda$, the IRS  can be modeled as a continuous surface with a continuous phase shift profile   denoted by $\Phi_{\text{irs}}(\mathbf{r})$ \cite{TCOM_IRSFSO}, where $\mathbf{r}=[x,y]^T$ denotes a point in the $xy$-plane and $(\cdot)^T$ refers to transposition.  The  phase shift introduced by the IRS is exploited such that it compensates the phase difference between the incident and the desired beam. Given the positions of the LS, $\left(d_{l }, \mathbf{\Psi}_{l }\right)$, and the  PD, $\left(d_{p},\mathbf{\Psi}_{p}\right)$,  we adopt a linear phase shift profile  as follows \cite{IRS_FSO_WCNC}
	\begin{IEEEeqnarray}{rll}
		\phi_\text{irs}(\mathbf{r})&=k\left(\Phi_{0}+\Phi_{x} x+\Phi_{y} y\right),\qquad\label{LP}
	\end{IEEEeqnarray}
	where  $k=\frac{2\pi}{\lambda}$ is the wave number, $\lambda$ is the wavelength, $\Phi_{x}=\cos(\theta_{l })\cos(\phi_{l})+\cos(\theta_{p})\cos(\phi_{p})$,  $\Phi_{y}=\cos(\theta_{l })\sin(\phi_{l })+\cos(\theta_{p})\sin(\phi_{p})$, and $\Phi_{0}$ is constant.  
	
	\subsubsection{Discrete-Time System Model}
	By sampling signal $u(t)$ at times $t=mT$ and defining $u[m]\overset{\Delta}{=} u(mT)$,  we obtain
	\begin{IEEEeqnarray}{rll}
		u[m]&{=} \sum\limits_{\ell=-\infty}^{\infty}h_e[\ell] {a}[m-\ell]+n[m]\nonumber\\
		&=h_e[0]{a}[m]+\sum_{\ell \neq m}^{}h_e[\ell]{a}[m-\ell]+n[m],
		\label{Discrete_model}
	\end{IEEEeqnarray}
	where  $h_e[m]\overset{\Delta}{=}h_e(mT)$ and $h_e(t)=h(t)\circledast g_\text{tx}(t)\circledast g_\text{rx}(t)$ is the overall impulse response. $n[m]=n(mT)$ has autocorrelation function $\Phi_{nn}[m]={N_0\over 2}\delta[m]$, i.e., the discrete-time noise is white. Here, $\delta[\cdot]$ denotes the Kronecker delta function. If the channel delay  dispersion is negligible and there is no ISI in the channel, then, $h_e[m]=h_\text{0} \delta[m]$, where $h_\text{0}$ is a constant. This is  the conventional system model for point-to-point non-dispersive FSO  channels. 
	
	\subsection{Channel Model}\label{Sec_Channel}
	IRS-assisted FSO channels are impaired by  geometric and misalignment losses (GML),  atmospheric loss, and  atmospheric turbulence induced fading \cite{IRS_FSO_WCNC}.  
	Thus, the  CIR  of the considered FSO system can be modeled as follows
	\begin{IEEEeqnarray}{rll}
		h(t)=R h_{p} h_{a} h_{\text{gml}}(t), \quad
		\label{channel}
	\end{IEEEeqnarray}
	where $R$ is the PD responsivity, $h_{a}$ represents the random atmospheric turbulence induced fading,  $h_{p}$ is the atmospheric loss, and  $h_{\text{gml}}(t)$ characterizes the GML impulse response. 
	
	In (\ref{channel}), to characterize the temporal characteristics of the IRS-assisted FSO channel, we ignore the  delay dispersion due to atmospheric turbulence. In fact, it was shown in \cite{Delayspread_turbulence} that for typical data rates of FSO links, which are on the order of several Gbps and a link distance of 1 km, the  root mean square (RMS) delay spread due to atmospheric turbulence is about 10 ps in rainy weather \cite{Delayspread_turbulence}. Given  symbol durations on the order of 0.1 ns, the delay dispersion due to atmospheric turbulence  is practically negligible \cite{Survey_Khalighi}.
	
	On the other hand,  the GML impulse response $h_{\text{gml}}(t)$ represents  the normalized  power collected at the lens disregarding  the atmospheric loss and atmospheric turbulence  and is given by  \cite{TCOM_IRSFSO}
	%
	\begin{IEEEeqnarray}{rll}
		h_{\text{gml}}(t)=\frac{1}{2\eta P_0}\int_{\mathcal{A}_\text{lens}}  \lvert E_\text{rx}(\mathbf{r}_p,t)\rvert^2 \mathrm{d}\mathbf{r}_p,
		\label{gml_basic}
	\end{IEEEeqnarray}  
	where  $E_\text{rx}(\mathbf{r}_p,t)$ is the received electric field at time $t$, $\eta$ is the free-space impedance,  $\mathcal{A}_\text{lens}$  denotes the area of the Rx lens,  and $\mathbf{r}_{p}=[x_{p}, y_{p}]^T$ denotes a   point in the lens plane.  Here, the origin of the $x_{p}y_{p}z_{p}$-coordinate system is  the center of the   lens and the $z_{p}$-axis  is parallel to  the normal vector of the  lens plane. We assume that the $y_{p}$-axis is parallel to the intersection line of the lens plane and the $xy$-plane, and the $x_{p}$-axis is perpendicular to the $y_{p}$- and $z_{p}$-axes.

	To determine  $E_\text{rx}(\mathbf{r}_p,t)$, we assume that the LS emits a Gaussian laser beam at time  $t-\tau$, which is reflected by the IRS and received at the lens with delay $\tau$. Then, employing the Huygens-Fresnel principle \cite{Goodmanbook}, we obtain
	\begin{IEEEeqnarray}{rll}
		&E_\text{rx}(\mathbf{r}_p,t)=\frac{\zeta}{\mathrm{j}\lambda}\times\nonumber\\
		&\iint\limits_{(x,y)\in\Sigma_\text{irs}} \frac{|E_\text{in}(\mathbf{r})|}{\lVert\mathbf{r}_o-\mathbf{r}\rVert} e^{-\mathrm{j}\left(k\lVert\mathbf{r}_o-\mathbf{r}\rVert+\phi_\text{irs}(\mathbf{r})+\psi_\text{in}(\mathbf{r})\right)} \delta\left(t-\tau(\mathbf{r})\right) \mathrm{d}\mathbf{r},	
		\label{Huygens_Delayed}
	\end{IEEEeqnarray}
	where  $\Sigma_\text{irs}$ is the IRS area, $\zeta$ is  the passivity factor  of the IRS,  and $|E_\text{in}(\mathbf{r})|$  and $\psi_\text{in}(\mathbf{r})$ are respectively the amplitude and phase of the incident  Gaussian beam  on the IRS.
	Moreover, $\delta(\cdot)$ is the Dirac delta function and $\tau(\mathbf{r})$ is the delay profile across the IRS. Furthermore,   $\mathbf{r}_o=[x_o,y_o,z_o]^T$ is the observation point on the lens plane in $xyz$-coordinates and is given by $\mathbf{r}_o=\mathbf{R}_\text{rot} [\mathbf{r}_p^T,d_p]^T$, where $\mathbf{R}_\text{rot}=\left(\begin{smallmatrix}
	\cos(\phi_p) \sin(\theta_p) &-\sin(\phi_p)&\cos(\phi_p) \cos(\theta_p)\\
	\sin(\phi_p) \sin(\theta_p) &\cos(\phi_p) &\sin(\phi_p) \cos(\theta_p)\\
	-\cos(\theta_p) &0 & \sin(\theta_p)\\
	\end{smallmatrix}\right)$.

	Adopting the paraxial approximation \cite{Andrews_book},  the Gaussian laser beam propagates along the beam axis and 
	the electric field incident on the IRS is then given by \cite{TCOM_IRSFSO}
	\begin{IEEEeqnarray}{rll}
		&E_\text{in}\!\left(\mathbf{r},d_l\right)= \frac{E_0 w_0}{w(d_l)} \exp\left({-{x^2\over w_x^2(d_l)}-{y^2\over w_y^2(d_l)}-\mathrm{j}\psi_\text{in}}(\mathbf{r})\right)\nonumber\\
		&\psi_\text{in}(\mathbf{r})=k \left(d_l-x\cos(\theta_l)+\frac{x^2}{2R_x(d_l)}+\frac{y^2}{2R_y(d_l)}\right)-\psi_o(d_l),\,
		\quad
		\label{Gauss-incident}
	\end{IEEEeqnarray}
	where $E_0=\sqrt{\frac{4\eta P_0}{\pi w_0^2}}$, $\psi_0(d_l)=\tan^{-1}\left(\frac{d_{l}}{z_{R}}\right)$,  
	$w(z_{l})=w_{0}\Big[{1+\left(\frac{z_{l}}{z_{R }}\right)^2}\Big]^{1/2}$ is the beamwidth at distance $z_{l}$, $w_{0}$ is the beam waist,  $z_{R }=\frac{\pi w_{0}^2}{\lambda}$ is the Rayleigh range, and $R(z_{l})=z_{l}\Big[1+\left(\frac{z_{R }}{z_{l}}\right)^2\Big]$ is the radius of the curvature of the beam's wavefront. Moreover, $w_{x}(d_l)=\frac{w(d_l)}{\sin(\theta_l)}$, $w_{y}(d_l)=w(d_l)$, $R_{x}(d_l)=\frac{R(d_l)}{\sin^2(\theta_l)}$, and $R_{y}(d_l)=R(d_l)$.

	%
	
	To solve (\ref{Huygens_Delayed}),  in the next section, we first derive the delay profile $\tau(\mathbf{r})$. Then, we employ scattering theory to determine the delayed received electric field and an analytical expression for the corresponding GML impulse response. 
	
	
	\section{Derivation of  CIR }
	To model the CIR, we first derive the delay considering  the distance between Tx and IRS, the distance between  Rx and IRS, and the orientation of  Tx and Rx with respect to (w.r.t.) the IRS. 
	\subsection{Delay Model}
	The end-to-end delay of the beam originating from the LS, reflected by the IRS, and received at the lens can be modeled by the optical path through which the beam propagates. Moreover, the end-to-end optical path is closely linked  to the total accumulated phase 
	from the Tx to the Rx via $L_\text{e2e}=\frac{\Phi_\text{tot}(\mathbf{r},\mathbf{r}_o)}{k}$, where $\Phi_\text{tot}(\mathbf{r},\mathbf{r}_o)=\psi_\text{in}(\mathbf{r})+\psi_\text{out}(\mathbf{r},\mathbf{r}_o)$. Here, $\psi_\text{out}(\mathbf{r},\mathbf{r}_o)=k\lVert\mathbf{r}_o-\mathbf{r}\rVert$ is the accumulated phase of the reflected beam to the lens and $\psi_\text{in}(\mathbf{r},\mathbf{r}_o)$ is the accumulated phase of the incident beam on the IRS \cite{TCOM_IRSFSO}. Thus, given the speed of light in  air, $v_l$, the end-to-end delay is given by 
	\begin{IEEEeqnarray}{rll}
		\tau(\mathbf{r},\mathbf{r}_o)=\frac{L_\text{e2e}}{v_l}=\frac{1}{kv_l}\left[\psi_\text{in}(\mathbf{r})+{\psi_\text{out}(\mathbf{r},\mathbf{r}_o)}\right].
		\label{rho}
	\end{IEEEeqnarray}
	The above equation shows that the end-to-end delay  depends on the phase profiles of the incident and reflected beams,   the considered points  on the IRS, $\mathbf{r}$, and the considered point on the lens, $\mathbf{r}_o$.

	Given that  the size of the  IRS and the width of the incident beam on the IRS are typically much  smaller than the IRS-to-Rx distance, the IRS typically operates in the intermediate regime (Fresnel regime) \cite{TCOM_IRSFSO}, where $d_p\gg d_f$, $d_f=\frac{x_e^2+y_e^2}{2\lambda}$, and $i_e=\min(\frac{L_i}{2},w_i(d_l)), i\in\{x, y\}$. For this case, it was shown in \cite{TCOM_IRSFSO}  that considering  only the first and second order terms of $\mathbf{r}$  in $\psi_\text{out}(\mathbf{r})$  is sufficient, which leads to the following approximation 
	\begin{IEEEeqnarray}{rll}
		\psi_\text{out}(\mathbf{r})=-k	\lVert\mathbf{r}_o-\mathbf{r}\rVert\approx -k\left(d_p-\mathbf{a}^T\mathbf{r}+\mathbf{r}^T\mathbf{R}_1\mathbf{r}\right),
		\label{fresnel-fraunhofer}
	\end{IEEEeqnarray}
	where $\mathbf{a}=\frac{1}{d_p}[x_o, y_o]^T$ and $\mathbf{R}_1=\frac{1}{2d_p}
	\begin{pmatrix}
	1-\frac{x_o^2}{d_p^2}&\frac{-x_oy_o}{d_p^2}\\
	\frac{-x_oy_o}{d_p^2}&1-\frac{y_o^2}{d_p^2}
	\end{pmatrix}$. Substituting (\ref{Gauss-incident}) and (\ref{fresnel-fraunhofer})  in (\ref{rho}), $\tau(\mathbf{r}, \mathbf{r}_o)$ is given by
	\begin{IEEEeqnarray}{rll}
		\tau(\mathbf{r}, \mathbf{r}_o)&\approx \tau_0+\mathbf{c}^T\mathbf{r}+\mathbf{r}^T \mathbf{B} \mathbf{r},
		\label{Tau_general}
	\end{IEEEeqnarray}
	where $\mathbf{c}= \frac{1}{v_l}\left(-\mathbf{a}-\mathbf{b}\right)$, $\mathbf{b}=[\cos(\theta_l)\cos(\phi_l), 0]^T$, $\mathbf{B}= \frac{1}{v_l}\left(\mathbf{R}_1+\mathbf{R}_2\right)$,  $\mathbf{R}_2=\begin{pmatrix}
	\frac{1}{2R_x(d_l)}&0\\
	0&\frac{1}{2R_y(d_l)}
	\end{pmatrix}$,  $\tau_0=\frac{-\psi_0}{v_l}+\tau_\text{los}$, and  $\tau_\text{los}= \frac{d_l+d_p}{v_l}$ is the end-to-end delay of the LOS link without  delay dispersion.
	Eq.~(\ref{Tau_general}) shows that the delay profile  is a function of $x^2$ and $y^2$, and thus,  the points on the IRS causing similar delay are located on an ellipse.
	
	We can further simplify (\ref{Tau_general}) by taking into account that the    lens   and  IRS sizes  are typically much smaller than the   Tx-to-IRS and IRS-to-Rx distances, i.e., ${a}\ll d_p$, and  $L_x, L_y\ll \min\{\sqrt{2d_l},\sqrt{2d_p}\}$. In this case,  the linear terms  of $\mathbf{r}$  in (\ref{Tau_general}) are  dominant and thus, by substituting  $\mathbf{r}_o=\mathbf{R}_\text{rot} [\mathbf{r}_p^T,d_p]^T$, the delay profile can be  simplified to a linear model as follows
	\begin{IEEEeqnarray}{rll}
		\tau(\mathbf{r})&\approx \tau_0+\mathbf{c}^T\mathbf{r}=\tau_0+a_1{x}+a_2{y},\qquad
		\label{Lineardelay_farfield}
	\end{IEEEeqnarray}
	where $a_1=\frac{-1}{v_l}\left[\cos(\phi_p)\cos(\theta_p)+\cos(\phi_l)\cos(\theta_l)\right]$ and $a_2=\frac{-1}{v_l}\left[\sin(\phi_l)\cos(\theta_l)+\sin(\phi_p)\cos(\theta_p)\right]$.
	Eq.~(\ref{Lineardelay_farfield}) reveals that the delay profile 
	depends on the positions and orientations of the LS and the PD, i.e., $(d_l,\Psi_{l})$ and $(d_p, \Psi_p)$. Defining the delay spread as the difference between the maximum and the minimum value of the delay profile, i.e., $\Delta\tau(\mathbf{r})=\max \left(\tau(\mathbf{r})\right)-\min\left(\tau(\mathbf{r})\right)$,  for IRS lengths $L_x=L_y=1$ m, $v_l=3\times 10^8\,\frac{\mathrm{m}}{\mathrm{s}}$,  $\Psi_{l }=(\frac{\pi}{2},0)$, and $\Psi_{p}=(0,\pi)$, we obtain a  delay spread of $3.3$ ns.  Given the small pulse duration for high data rate FSO systems,  e.g., $T=0.1$ ns,  the delay spread is expected to be much  larger than the pulse width, i.e., $\Delta\tau(\mathbf{r})\gg T$, thus, successively transmitted pulses overlap at the Rx  causing  ISI. On the other hand, if $\Delta\tau(\mathbf{r})\ll T$ holds for the delay spread, the  waves reflected by different  IRS elements  result in constructive or destructive interference and no pulse dispersion is expected. For the mentioned example, the delay spread can be ignored if  $\cos(\theta_{p })-\cos(\theta_l)\ll \frac{Tv_l}{L_x}=0.03$, and given $\theta_l=\frac{\pi}{2}$, we obtain $\theta_p\gg 1.54\,\mathrm{rad}\approx \frac{\pi}{2}$. This example shows that the delay spread is negligible when the elevation angles of the Tx and Rx w.r.t. the IRS are identical  and Tx and Rx are located in a  plane perpendicular to the IRS, i.e.,   $\theta_l\approx\theta_p$, $\phi_l=0$, and  $\phi_p=\pi$.
	Next,  using the linear delay profile in (\ref{Lineardelay_farfield}), we derive the dispersive CIR in the next section.
	
	\subsection{ Channel Impulse Response}
	To determine the impact of delay dispersion on the optical electric field, we  employ a similar approach as the Knife-edge diffraction model \cite{Goldsmith_book, Barclay_KnifeEdge}, where we consider the different portions of the beam  incident  on the IRS as Huygens secondary sources and the sum of these sources determines  the received wavefront on the lens. The electric field emitted by each of  the secondary sources arrives at the lens with an end-to-end delay of $\tau(\mathbf{r})$, and  the resulting  received field is obtained based on (\ref{Huygens_Delayed}). In the following lemma, we simplify (\ref{Huygens_Delayed}) using the delay profile in (\ref{Lineardelay_farfield}).
	
	\begin{lem}
		Assuming the linear delay profile in (\ref{Lineardelay_farfield}), the dispersive electric field received at the lens at time $t$ is given by
		\begin{IEEEeqnarray}{rll}
			&E_\text{rx}(\mathbf{r}_o,t){=}\int_{\varkappa\in\mathcal{C}} E(\mathbf{r}(\varkappa))e^{-\mathrm{j}\phi(\mathbf{r}(\varkappa), \mathbf{r}_o)} \frac{\lVert\nabla\mathbf{r}(\varkappa)\rVert}{\lVert\nabla{\tau}({\mathbf{r}})\rVert}\,  \mathrm{d}\varkappa,
			\label{Cont-model_Erx}
		\end{IEEEeqnarray}	
		where $E(\mathbf{r}(\varkappa))=\frac{\zeta}{\mathrm{j}\lambda d_p}\left|E_\text{in}(\mathbf{r}(\varkappa))\right|$, $\phi(\mathbf{r}(\varkappa), \mathbf{r}_o)=k\lVert\mathbf{r}_o-\mathbf{r}(\varkappa)\rVert+\phi_\text{irs}(\mathbf{r}(\varkappa))+\psi_\text{in}(\mathbf{r}(\varkappa))$, $\nabla \mathbf{r}(\varkappa)$ denotes the gradient of vector $\mathbf{r}$ w.r.t. variable $\varkappa$, and $\mathcal{C}$ denotes the curve where $\tau(\mathbf{r}(\varkappa))=t$ holds.
	\end{lem}
	\begin{IEEEproof}
		First, we substitute  (\ref{Lineardelay_farfield}) in  (\ref{Huygens_Delayed}). Next,  to simplify the $2$-dimensional integral, we use $\int_{\mathcal{R}^n}f(\mathbf{x}) \delta\left(g(\mathbf{x})\right)\mathrm{d}\mathbf{x}=\int_{g^{-1}(0)}\frac{f(\varkappa)}{\lvert\nabla g(\mathbf{x})\rvert} \mathrm{d}\sigma(\varkappa)$, which is known as the ``simple layer integral'' and  $g^{-1}(0)$ denotes the $(n-1)$-dimensional surface where $g(\mathbf{x})=0$ \cite{Dirac_book}. In (\ref{Huygens_Delayed}), we have $\mathbf{x}=\mathbf{r}$, $g(\mathbf{x})=t-\tau(\mathbf{r})$,  and  $f(\mathbf{x})=E(\mathbf{r}) e^{-\mathrm{j}\phi(\mathbf{r})} $  in the simple layer integral. Then, assuming   $\mathbf{r}(\varkappa)=[x(\varkappa),y(\varkappa)]^T$, and substituting $\mathrm{d}\sigma(\varkappa)$ by $\lVert\nabla\mathbf{r}(\varkappa)\rVert\mathrm{d}\varkappa$, we obtain (\ref{Cont-model_Erx}) and this completes the proof.
	\end{IEEEproof}

	\begin{theo}\label{Theo1}
		The CIR of an IRS-assisted FSO link  can be accurately approximated by
		\iftoggle{Conf}{%
				\begin{IEEEeqnarray}{rll}
				&h_\text{gml}(t)=C_he^{-\sigma_\tau\left({t-\tau_0}\right)^2}\times\nonumber\\
				&\int\limits_{-\tilde{a}}^{\tilde{a}}\int\limits_{-\tilde{a}}^{\tilde{a}}e^{\mathbf{r}_p^T\mathbf{D}^T\mathbf{C}\mathbf{D}\mathbf{r}_p+\frac{t-\tau_0}{a_2}\mathbf{e}\mathbf{D}\mathbf{r}_p}
				\mathrm{d}\mathbf{r}_p,\,\,  \left|t-\tau_0\right|\leq a_1\frac{L_x}{2}+a_2\frac{L_y}{2},\qquad
				\label{Cont_model}
			\end{IEEEeqnarray}
		}{%
				\begin{IEEEeqnarray}{rll}
				h_\text{gml}(t)&=C_he^{-\sigma_\tau\left({t-\tau_0}\right)^2}\int\limits_{-\tilde{a}}^{\tilde{a}}\int\limits_{-\tilde{a}}^{\tilde{a}}e^{\mathbf{r}_p^T\mathbf{D}^T\mathbf{C}\mathbf{D}\mathbf{r}_p+\frac{t-\tau_0}{a_2}\mathbf{e}\mathbf{D}\mathbf{r}_p}
				\mathrm{d}\mathbf{r}_p,\quad  \left|t-\tau_0\right|\leq a_1\frac{L_x}{2}+a_2\frac{L_y}{2},
				\label{Cont_model}
			\end{IEEEeqnarray}
		}
		where $\sigma_\tau=2\frac{c_2}{a_2^2}-c_2^2\frac{c_5-2\frac{a_1}{a_2}}{2a_2^2a_s}$, $\tilde{a}={\sqrt{\pi}a\over 2}$, $C_h=\frac{2\zeta^2}{\lambda^2d_p^2|a_2|^2|a_s|w^2(d_l)}$, $\mathbf{D}=\frac{-\mathrm{j}k}{2d_p}\begin{pmatrix}
		\cos(\phi_p) \sin(\theta_p) &-\sin(\phi_p)\\
		\sin(\phi_p) \sin(\theta_p) &\cos(\phi_p) \\
		\end{pmatrix}$, $\mathbf{C}=\frac{1}{2a_s}\begin{pmatrix}
		1  &-\frac{a_1}{a_2}\\
		\frac{a_1}{a_2} &1 \\
		\end{pmatrix}$,  and $\mathbf{e}=\left[\frac{1}{a_s}, -2-\frac{c_0a_1}{a_sa_2} \right]$. Here,  $c_0=c_2\left(c_5-2\frac{a_1}{a_2}\right)$, $c_1=\frac{1}{w_x^2(d_l)}+\frac{\mathrm{j}k}{2d_p}\left(1-\cos^2(\phi_p)\cos^2(\theta_p)\right)+\frac{\mathrm{j}k}{2R_x(d_l)}$, $c_2=\frac{1}{w_y^2(d_l)}+\frac{\mathrm{j}k}{2d_p}\left(1-\sin^2(\phi_p)\cos^2(\theta_p)\right)+\frac{\mathrm{j}k}{2R_y(d_l)}$,  $c_5= \frac{-\mathrm{j}k}{d_p}\sin(\phi_p)\cos(\phi_p)\cos^2(\theta_p)$, and $a_s=c_1+\left(\frac{a_1}{a_2}\right)^2c_2-\frac{a_1}{a_2}c_5$.
	\end{theo}
	\begin{IEEEproof}
		The proof is provided in Appendix \ref{App1}.
	\end{IEEEproof}
	Based on (\ref{Cont_model}), $h_\text{gml}(t)$ can be evaluated numerically. As the integral limits are finite, the complexity of numerical integration is low.
	\begin{corol} \label{Cor1}
		If the LS and PD  are located  at $\phi_l=0$ and $\phi_p=\pi$ (in-plane reflection), respectively, the linear delay profile in (\ref{Lineardelay_farfield}) simplifies to $\tau(\mathbf{r})=a_1x+\tau_0$. Thus, the CIR  and  its frequency response can be respectively  accurately approximated by
		\iftoggle{Conf}{%
			\begin{IEEEeqnarray}{rll}
				&h_\text{gml}(t)=\begin{cases}
					\sqrt{c_\tau\over \pi}h_\text{LOS}\exp\left(-c_\tau(t-\tau_0)^2\right),&\left|t-\tau_0\right| \leq \frac{L_x}{2}a_1,\\
					0, &\text{otherwise},
				\end{cases}
				\nonumber\\
				&H_\text{gml}(f)= \frac{-1}{2}h_\text{LOS} e^{-\pi^2 f^2/ c_\tau}\nonumber\\
				&\times\left[\text{erf}\left(\frac{\sqrt{c_\tau}}{2}L_x a_1-\mathrm{j}\frac{\pi f}{\sqrt{c_\tau}}\right)+\text{erf}\left(\frac{\sqrt{c_\tau}}{2}L_x a_1+\mathrm{j}\frac{\pi f}{\sqrt{c_\tau}}\right)\right]\!,
				\label{Corol}
			\end{IEEEeqnarray}
		}{%
			\begin{IEEEeqnarray}{rll}
				&h_\text{gml}(t)=\begin{cases}
					\sqrt{c_\tau\over \pi}h_\text{LOS}\exp\left(-c_\tau(t-\tau_0)^2\right),&\left|t-\tau_0\right| \leq \frac{L_x}{2}a_1,\\
					0, &\text{otherwise},
				\end{cases}
				\nonumber\\
				&H_\text{gml}(f)= \frac{-1}{2}h_\text{LOS} e^{-\pi^2 f^2/ c_\tau}\left[\text{erf}\left(\frac{\sqrt{c_\tau}}{2}L_x a_1-\mathrm{j}\frac{\pi f}{\sqrt{c_\tau}}\right)+\text{erf}\left(\frac{\sqrt{c_\tau}}{2}L_x a_1+\mathrm{j}\frac{\pi f}{\sqrt{c_\tau}}\right)\right],
				\label{Corol}
			\end{IEEEeqnarray}
		}
		where  $h_\text{LOS}=\text{erf}\left(\frac{k\tilde{a}}{\sqrt{2}w(d_l)|b_y|d_p}\right)$, $b_x= \frac{1}{w_x^2(d_l)}+\frac{\mathrm{j}k}{2R_x(d_l)}+\frac{\mathrm{j}k}{2d_p}$, $b_y= \frac{1}{w_y^2(d_l)}+\frac{\mathrm{j}k}{2R_y(d_l)}+\frac{\mathrm{j}k}{2d_p}$, and $c_\tau=\frac{2 }{a_1^2w_x^2(d_l)}$.
	\end{corol}
	\begin{IEEEproof}
		The proof is provided in Appendix \ref{App2}.
	\end{IEEEproof}
	
	As can be observed in (\ref{Corol}), the CIR is a truncated Gaussian function with parameter $c_\tau$ which depends on the width of the  beam footprint on the IRS, and the position of the LS and PD via $a_1=\frac{1}{v_l}\left[\cos(\theta_p)-\cos(\theta_l)\right]$. This reveals that for  $\theta_l=\theta_p$ the end-to-end channel is non-dispersive and by increasing the difference between angles $\theta_l$ and $\theta_p$, the maximum delay spread increases which leads to dispersion and ISI in such systems.
	
	\subsection{Conservation of Energy}
	Disregarding the path loss and channel fading for the moment,  the law of conservation of energy implies that   the power of   the beam incident on the IRS  is  equal to the  power reflected from the IRS. We take this  into account by factor $\zeta$. Assuming that the LS  transmits with average  power $P_0$,    the average received power at the lens is  given by
	\begin{IEEEeqnarray}{rll}
		\bar{{P}}_\text{rx}=P_0	\int_{-\infty}^\infty h_\text{gml}(\tau)\mathrm{d}\tau.
		\label{inst_power}
	\end{IEEEeqnarray}
	Given that the average received power is equal to the average transmit power due to the passivity of the IRS, the following proposition provides the value of $\zeta$ for in-plane IRS reflections.
	
	\begin{prop}\label{Prop1}
		For in-plane reflection, the passivity factor $\zeta$  of the IRS-based FSO channel in (\ref{Corol}) is given by 
		\begin{IEEEeqnarray}{rll}
			\zeta^2= {\lambda |a_1| d_p \sin(\theta_l)}.
			\label{zeta_Col}
		\end{IEEEeqnarray}
	\end{prop}
	\begin{IEEEproof}
		The proof is provided in Appendix \ref{App3}.
	\end{IEEEproof}
	\section{Simulation Results}
	
	\begin{figure}[t]
		\centering
		\includegraphics[width=0.65\textwidth]{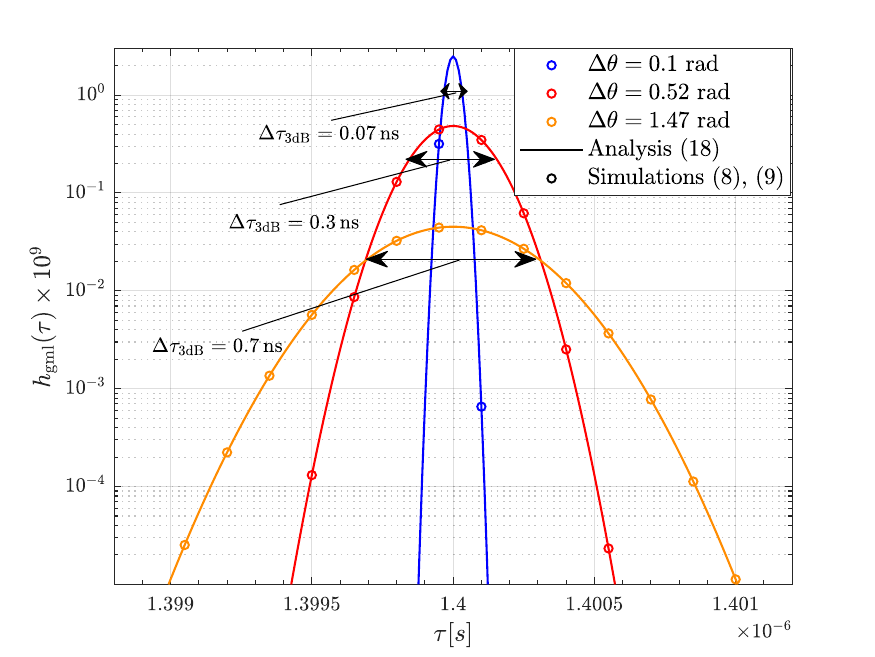}
		\vspace{-4mm}
		\caption { Channel impulse response versus delay for $\theta_l=\frac{\pi}{2}$ and different $\Delta\theta$.\label{Fig:GML}}
		\vspace{-4mm}
	\end{figure}

	In this section, first we validate the analytical CIR  in (\ref{gml_basic}) for a dispersive  IRS-assisted FSO link.  Then, we  investigate the system performance in terms of BER for different equalization methods.   We assume the LS emits a Gaussian beam with $w_0=1$ mm, $\lambda=1550$ nm, FSO bandwidth $W_\text{fso}=10$ GHz, $v_l=3\times 10^8\,{\mathrm{m} \over \mathrm{s}}$, and $P_0=0.05\, \mathrm{mW}$. The Rx is equipped with a circular lens with radius $a=10$ cm, and the  noise spectral density is $N_0=-104\ \mathrm{dBm}/\mathrm{MHz}$. Moreover, the Tx and Rx are  located at $(d_l,\mathbf{\Psi}_l)=(200\, \text{m}, [\frac{\pi}{2},0])$ and $(d_p,\mathbf{\Psi}_p)=(220\, \text{m}, [\theta_p,\pi])$, respectively.

	Fig.~\ref{Fig:GML} shows the CIR as a function of time for different values of $\Delta\theta=\theta_l-\theta_p$. We define the  \textit{effective delay spread} as the delay spread where the amplitude of the CIR $h_\text{gml}(\tau)$ reduces by 3 dB compared to  its maximum value, which is denoted  by $\Delta\tau_{3\, \mathrm{dB}}$. As can be observed, by decreasing the value of $\Delta\theta$, the  delay  reduces to  the LOS end-to-end delay of $\tau_\text{los}=1.4\,\mu \mathrm{s}$.  Here, for a Rx located at $\theta_p=0.1$ rad ($\Delta\theta=1.47$ rad),  the FSO link experiences an effective  delay spread of $0.7$ ns, whereas for  Rxs at angles $\theta_p=1.05$ rad ($\Delta\theta=0.52$ rad) and $\theta_p=1.47$ rad ($\Delta\theta=0.1$ rad), the  effective delay spreads reduce to 0.3 ns and 0.07 ns, respectively.   We can observe that by increasing the difference between the angle of the incident beam $\theta_l$ and the angle of the reflected beam $\theta_p$, i.e., $\Delta\theta$, the effective delay spread induced by the IRS increases. This can be seen in the delay model in (\ref{Lineardelay_farfield}) from parameter $a_1=\frac{1}{v_l}\left[\cos(\theta_p)-\cos(\theta_l)\right]$. Fig.~\ref{Fig:GML} also shows that by increasing  angle $\theta_p$ (decreasing $\Delta\theta$), the CIR  has larger maximum values and thus,  the  lens receives more power from the IRS. This means that, for smaller $\Delta\theta$, the channel is less dispersive and the geometric loss is smaller.
		
		\begin{figure}[t]
			\centering
			\includegraphics[width=0.65\textwidth]{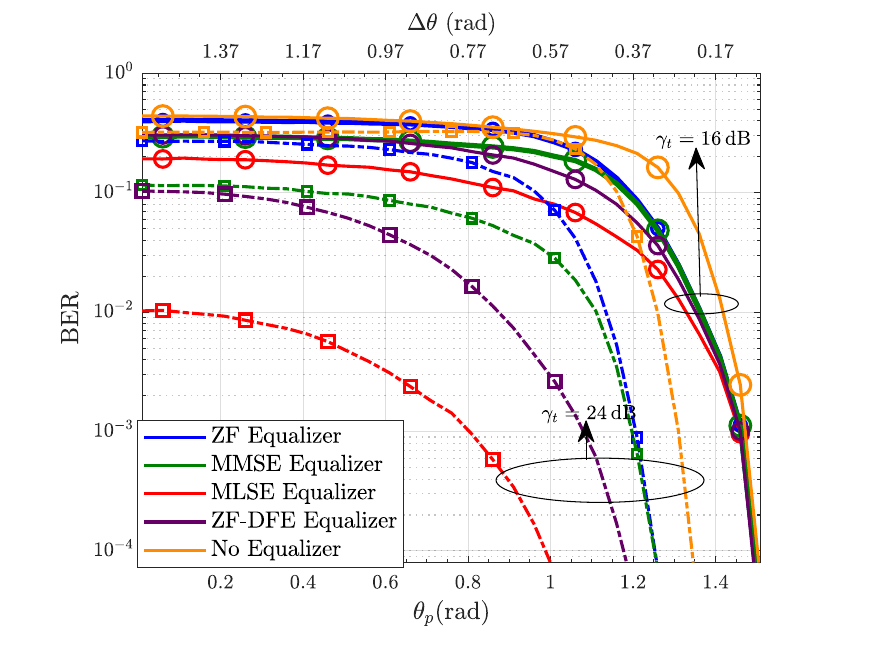}
			\vspace{-4mm}
			\caption { BER versus angles $\theta_p$ and $\Delta\theta$.\label{Fig:BER}}
			\vspace{-4mm}
		\end{figure}

	Fig.~\ref{Fig:BER} shows the BER versus angles $\theta_p$ and $\Delta\theta=\frac{\pi}{2}-\theta_p$ for   transmit SNRs $\gamma_t= 16$ dB and 24 dB. To alleviate the ISI caused by delay spread, we employ   maximum likelihood sequence estimation (MLSE), linear zero forcing (ZF) and  minimum mean squared error (MMSE) equalization, and ZF-decision feedback equalization (ZF-DFE) at the Rx \cite{Proakis}. By increasing $\theta_p$ from 0 rad to $\pi\over 2$ rad (decreasing $\Delta\theta$), the number of dispersive FSO channel  taps gradually decrease from 10 taps  to 1 tap. Thus, we implemented  MLSE  using the Viterbi algorithm with  $2^{9}$ states and linear equalizers  with 20 taps. As can be observed, as $\theta_p$ increases, the BER decreases (even without  equalizer)   because of the smaller effective delay spread  of the IRS-based FSO channel and the larger GML gain due to the higher received power at the lens. Our results also show that   MLSE  yields the best performance 
	at the expense of higher complexity. Using linear equalization  techniques, we can reduce the complexity compared to Viterbi algorithm. In particular, the MMSE equalizer performs  better than the ZF equalizer for $\theta_p\leq 1.2$ rad ($\Delta\theta\geq0.37$ rad), since the MMSE equalizer reduces noise enhancement. On the other hand,  for $\theta_p>1.2$ rad ($\Delta\theta<0.37$ rad), the BER performance of the ZF equalizer  matches that of the MMSE equalizer. Moreover, for  $\gamma_t=16$ dB, both linear  equalizers approach the performance of MLSE for $\theta_p>1.4$ rad ($\Delta\theta< 0.17$ rad). Here, the difference between incident  and reflected angles is very small, and as a result, the delay spread and ISI  are also small. 
	Moreover, ZF-DFE provides a good compromise between performance and complexity. It outperforms both linear equalizers but requires a much lower complexity compared to MLSE \cite{Proakis}. 
	
	\section{Conclusions}
	In this paper, we investigated the delay dispersion introduced by an IRS  in an FSO link. Using our analysis, we developed  an analytical model for the delay profile and the continuous CIR of IRS-assisted FSO links. We  showed that equalization techniques can be used to overcome  the IRS-induced ISI. Our simulation results show that the positions of  Tx and Rx w.r.t. the IRS, the IRS size, and the incident beamwidth affect the maximum delay spread and link performance. Moreover, using suitable equalization techniques the IRS-induced ISI can be resolved at the Rx.
	\appendices
	\renewcommand{\thesectiondis}[2]{\Alph{section}:}
	\vspace*{-3mm}
	\section{Proof of Theorem~\ref{Theo1}}\label{App1}
	Considering the received electric field in (\ref{Cont-model_Erx}), and  given  $\tau(\mathbf{r})=\tau_0+a_1x+a_2 y$, we obtain $\lVert\nabla{\tau}({\mathbf{r}})\rVert=\lVert[a_1,a_2]^T\rVert=\sqrt{a_1^2+a_2^2}$. Then, using  $\mathbf{r}(s)=[x(s),y(s)]^T=[s,\tilde{\tau}_1-\frac{a_1}{a_2}s]$, with $\tilde{\tau}_1=\frac{t-\tau_0}{a_2}$ and $-\frac{L_x}{2}\leq s\leq \frac{L_x}{2}$, we have $\lVert\nabla\mathbf{r}(s)\rVert=\lVert[1,-\frac{a_1}{a_2}]\rVert=\left[1+\left(\frac{a_1}{a_2}\right)^2\right]^{1\over 2}$.
	Thus, by substituting (\ref{Gauss-incident}) and (\ref{LP}),  we obtain 
	\begin{IEEEeqnarray}{rll}
		&E_\text{rx}(\mathbf{r}_o,t)=\frac{E_0w_0\zeta}{\mathrm{j}\lambda d_p|a_2|w(d_l)} e^{-\mathrm{j}k(d_l+d_p)+\mathrm{j}\psi_0-\mathrm{j}k\Phi_0}
		e^{-c_2\tilde{\tau}_1^2-c_4\tilde{\tau}_1}\nonumber\\
		&\times\int_{-\frac{L_x}{2}}^{\frac{L_x}{2}}e^{-a_ss^2-b_ss}\mathrm{d}s\nonumber\\
		&\overset{(i)}{=}\frac{E_0w_0\zeta}{2\lambda d_p|a_2|w(d_l)}\sqrt{\frac{\pi}{a_s}} e^{-\mathrm{j}k(d_l+d_p)+\mathrm{j}\psi_0}
		e^{-c_2\tilde{\tau}_1^2-c_4\tilde{\tau}_1}e^{\frac{b_s^2}{4a_s}}\nonumber\\
		&\times\left[\text{erf}\left(\frac{\sqrt{a_s}L_x}{2}+\frac{b_s}{2\sqrt{a_s}}\right)-\text{erf}
		\left(-\frac{\sqrt{a_s}L_x}{2}+\frac{b_s}{2\sqrt{a_s}}\right)\right],\qquad
		\label{Proof2_Cont_model}
	\end{IEEEeqnarray}
	where in $(i)$ we use \cite[(2.33-1)]{integral}.  Moreover, $\bar{c}_1=\frac{1}{w_x^2(d_l)}+\frac{\mathrm{j}k}{2d_p}\left(1-\frac{x_o^2}{d_p^2}\right)+\frac{\mathrm{j}k}{2R_x(d_l)}$, $\bar{c}_2=\frac{1}{w_y^2(d_l)}+\frac{\mathrm{j}k}{2d_p}\left(1-\frac{y_o^2}{d_p^2}\right)+\frac{\mathrm{j}k}{2R_y(d_l)}$,  $c_3=\frac{-\mathrm{j}kx_o}{d_p}+\mathrm{j}k\Phi_x-\mathrm{j}k\cos(\theta_l)$, $c_4=\frac{-\mathrm{j}ky_o}{d_p}+\mathrm{j}k\Phi_y$, $\bar{c}_5=-\mathrm{j}k\frac{x_oy_o}{d_p^3}$, and $b_s=c_3-\frac{a_1}{a_2}c_4+\tilde{\tau}_1(c_5-2\frac{a_1}{a_2})c_2$.  
	
	Then, assuming $L_x\gg {2\over \sqrt{a_s}}$, we obtain the power density, defined as $I_\text{rx}(\mathbf{r}_o,t)=\frac{1}{2\eta P_0} \left|E_\text{rx}(\mathbf{r}_o,t)\right|^2$, as follows
	\begin{IEEEeqnarray}{rll}
		I_\text{rx}(\mathbf{r}_o,t)&=C_h\exp\left(\frac{1}{2a_s}\left(c_3-c_4\frac{a_1}{a_2}\right)^2\right)\nonumber\\
		&\times\exp\Biggl(-\tilde{\tau}_1^2\left(2c_2-\left(c_5-2\frac{a_1}{a_2}\right)^2\frac{c_2^2}{2a_s}\right)\nonumber\\
		&-\tilde{\tau}_1\left(2c_4-\frac{c_2}{a_s}\left(c_5-2\frac{a_1}{a_2}\right)\left(c_3-\frac{a_1}{a_2}c_4\right)\right)\Biggl).\qquad
		\label{Proof3_Cont_model}
	\end{IEEEeqnarray}
	Then, we apply  $\mathbf{r}_o=\mathbf{R}_3\left[\mathbf{r}_p^T,d_p\right]^T$ and approximate $\frac{x_o^2}{d_p^2}\approx\cos^2(\phi_p)\cos^2(\theta_p)$,  $\frac{y_o^2}{d_p^2}\approx\sin^2(\phi_p)\cos^2(\theta_p)$, and $\frac{x_o y_o }{d_p^2}\approx \sin(\phi_p)\cos(\phi_p)\cos^2(\theta_p)$ \cite{TCOM_IRSFSO}, which leads to $\bar{c}_1\approx c_1$, $\bar{c}_2\approx c_2$, and $\bar{c}_5\approx c_5$. Then, by applying the values of $\Phi_x$ and $\Phi_y$ in (\ref{LP}) and substituting  $[c_3,c_4]^T=\mathbf{D} \mathbf{r}_p$ and (\ref{Proof3_Cont_model}) in (\ref{gml_basic}), we obtain (\ref{Cont_model}). This completes the proof.

	
	\section{Proof of Corollary~\ref{Cor1}}\label{App2}
	Given (\ref{Cont-model_Erx}) and $\tau(\mathbf{r})=a_1 x+\tau_0$, we can substitute $\mathbf{r}(s)=[\tilde{\tau}_2, s]^T$ with $\tilde{\tau}_2=\frac{t-\tau_0}{a_1}$ and $-\frac{L_y}{2}\leq s\leq \frac{L_y}{2}$. Using similar approximations as in the proof of Theorem \ref{Theo1},  the received electric field at the lens obtained as
	\begin{IEEEeqnarray}{rll}
		&E_\text{rx}(\mathbf{r}_o, t)=\frac{1}{|a_1|}\int_{-\frac{L_y}{2}}^{\frac{L_y}{2}}E(\mathbf{r}(s)) e^{-\mathrm{j}\phi(\mathbf{r}(s),\mathbf{r}_o)} \mathrm{d}s\nonumber\\
		&=	 C_e e^{-\mathrm{j}k\phi_e+\mathrm{j}\psi_0} \exp\left(-\frac{(\tilde{\tau}_2)^2}{w_x^2(d_l)}\right)\int_{-\frac{L_y}{2}}^{\frac{L_y}{2}} e^{-b_ys^2-y(\mathrm{j}k\frac{y_p}{d_p})}\mathrm{d}s\nonumber\\
		&= C_e e^{-\mathrm{j}k\phi_e+\mathrm{j}\psi_0}\frac{\sqrt{\pi}}{2\sqrt{b_y}}\exp\left(-\frac{(\tilde{\tau}_2)^2}{w_x^2(d_l)}\right) \exp\left(\frac{-k^2y_p^2}{4b_yd_p^2}\right)\times\nonumber\\
		&\left[\text{erf}\left(\!\!\sqrt{b_y}\frac{L_y}{2}+\frac{\mathrm{j}ky_p}{2d_p\sqrt{b_y}}\!\right)-\text{erf}\left(\!\!-\sqrt{b_y}\frac{L_y}{2}+\frac{\mathrm{j}ky_p}{2d_p\sqrt{b_y}}\!\right)\!\right]\!,\quad
		\label{proof1-gml-simple}
	\end{IEEEeqnarray}
	where 
	$C_e=\frac{v_l\zeta E_0w_0}{\lambda d_p|a|w(d_l)}$, $\phi_e=d_l+d_p-\tilde{\tau} (\cos(\theta_l)-\cos(\theta_p))+\frac{(\tilde{\tau}_2)^2}{2R_x(d_l)}-{\tilde{\tau}_2}\frac{\sin(\theta_p)x_p}{d_p}+\left(\tilde{\tau}_2\right)^2\frac{\sin^2(\theta_p)}{2d_p}+\Phi_x\tilde{\tau}_2$.
	
	Assuming $L_y\gg \frac{2}{\sqrt{b_y}}$, we can simplify the received beam intensity as follows 
	\begin{IEEEeqnarray}{rll}
		I_\text{rx}(\tau,\mathbf{r}_p)=\frac{C_e^2 \pi}{2\eta P_0|b_y|}e^{-2\frac{\tilde{\tau}_2^2}{w_x^2(d_l)}} \exp\left(\frac{-k^2y_p^2\mathcal{R}\{b_y\}}{2|b_y|^2d_p^2}\right),\qquad
		\label{proof2-gml-simple}
	\end{IEEEeqnarray}
	where $\mathcal{R}\{\cdot\}$ denotes the real part of a complex number.
	Next, we obtain the GML factor in (\ref{gml_basic}) as follows
	\begin{IEEEeqnarray}{rll}
		h_\text{gml}(\tau)&=\frac{C_e^2 \pi}{2\eta P_0|b_y|}e^{-\frac{2(\tilde{\tau}_2)^2}{w_x^2(d_l)}} \iint_{\Sigma_\text{lens}}\exp\left(\frac{-k^2y_p^2\mathcal{R}\{b_y\}}{2|b_y|^2d_p^2}\right)\mathrm{d}x_p\mathrm{d}y_p\nonumber\\
		&\overset{(i)}{=}C_h\exp\left(-\frac{2\tilde{\tau}_2^2}{w_x^2(d_l)}\right) \text{erf}\left(\frac{k\tilde{a}}{\sqrt{2}w(d_l)|b_y|d_p}\right),
		\label{proof3-gml-simple}
	\end{IEEEeqnarray}
	where in $(i)$, we use $\int e^{ax^2}\mathrm{d}x=\frac{1}{2}\sqrt{\frac{\pi}{a}}\text{erfi}(\sqrt{a}x)$ and $\text{erfi}(x)=-\mathrm{j}\text{erf}(\mathrm{j}x)$. Moreover, $C_h=\sqrt{\frac{2}{\pi}}\frac{\zeta^2}{\lambda d_p|a_1|^2w(d_l)}$. Substituting $\zeta$ from (\ref{zeta_Col}), we obtain (\ref{Corol}) and this completes the proof.

	\section{Proof of Proposition~\ref{Prop1}}\label{App3}
	Ignoring the impact of atmospheric loss and atmospheric turbulence and given the passivity of the IRS, the average received power over the dispersive channel  should be equal to average transmitted power without delay given by
	\begin{IEEEeqnarray}{rll}
		\bar{P}_\text{rx}&=\int_{-\infty}^\infty	P_0 h_\text{gml}(t) \mathrm{d}t\nonumber\\
		&=P_0 	C_h \text{erf}\left(\frac{k\tilde{a}}{\sqrt{2}w(d_l)|b_y|d_p}\right)
		\int_{-\infty}^\infty	\exp\left(-\frac{2(\tilde{\tau}_2)^2}{w_x^2(d_l)}\right) \mathrm{d}t\nonumber\\	
		&\overset{(i)}{=}P_0\frac{\zeta^2}{\lambda d_p |a_1|\sin(\theta_l)} \text{erf}\left(\frac{k\tilde{a}}{\sqrt{2}w(d_l)|b_y|d_p}\right),
		\label{avg-power}
	\end{IEEEeqnarray}
	where in $(i)$ we use $\int_{-\infty}^{\infty} \exp(-p^2 x^2-qx)\,\mathrm{d}x=\frac{\sqrt{\pi}}{p}\exp\left(\frac{q^2}{4p^2}\right)$ \cite[(3.323-2)]{integral}. Comparing (\ref{avg-power}) with average transmit power $P_0$ leads to  (\ref{zeta_Col}) and completes the proof.

	\bibliographystyle{IEEEtran}
	\bibliography{My_Citation_1-07-2022}
\end{document}